\documentclass[prb,showpacs,twocolumn,preprintnumbers,amsmath,amssymb,floatfix]{revtex4}

\usepackage{graphicx,color}

\begin{document}

\title{Quantum Monte Carlo calculation of the Fermi liquid parameters of the
  two-dimensional homogeneous electron gas}

\author{N.\ D.\ Drummond}

\affiliation{Department of Physics, Lancaster University, Lancaster LA1 4YB,
  United Kingdom}

\author{R.\ J.\ Needs}

\affiliation{TCM Group, Cavendish Laboratory, University of Cambridge,
  J.\ J.\ Thomson Avenue, Cambridge CB3 0HE, United Kingdom}

\date{\today}

\begin{abstract}
Fermi liquid theory is the basic paradigm within which we understand the
normal behavior of interacting electron systems, but quantitative values for
the parameters that occur in this theory are currently unknown in many
important cases. One such case is the two-dimensional homogeneous electron gas
(2D HEG), which is realized in a wide variety of semiconductor devices.  We
have used quantum Monte Carlo (QMC) methods to calculate the Landau
interaction functions between pairs of quasiparticles.  We use these to study
the Fermi liquid parameters, finding that finite-size effects represent a
serious obstacle to the direct determination of Fermi liquid parameters in QMC
calculations. We have used QMC data in the literature for other properties of
the 2D HEG to assemble a set of ``best available'' values for the Fermi liquid
parameters.
\end{abstract}


\maketitle

\section{Introduction}

Many of the key theoretical developments in condensed matter physics have been
concerned with the exploration of models that capture important aspects of the
behavior of real materials.  One of the most fundamental and useful model
systems in the field is the \textit{homogeneous electron gas}
(HEG)\@.\cite{giuliani} The simplicity of the system (a gas of electrons
moving in a uniform, neutralizing background) is deceptive: the model exhibits
a rich range of physics and remains our basic starting point for understanding
the behavior of charge carriers in metals and semiconductors.

The enormous theoretical challenge that must be overcome when trying to
provide an accurate description of the HEG is that the electrons are strongly
coupled by their mutual Coulomb repulsion.  Nevertheless, many thermal and
transport properties of the HEG can be described by ignoring electron-electron
interactions altogether, resulting in the free-electron-gas model, in which
each electron has its own well-defined energy and momentum.  This observation,
which predates quantum mechanics, was first explained within a general
theoretical framework by Landau through the development of Fermi liquid
theory.\cite{landau} Although the existence of electron-electron repulsion and
hence correlation dramatically changes the total energy of an electron gas,
low-lying excitations have a nonvanishing overlap with the corresponding
excitations of the noninteracting system, in which the single-particle
orbitals are plane-wave momentum eigenstates. Hence we may associate each
excited state of the interacting system with a particular set of
\textit{quasiparticle} momentum occupation numbers.

Remarkably, although Fermi liquid theory is our basic paradigm for the normal
behavior of the fluid phase of an electron gas, quantitative values for the
parameters that occur in this theory are essentially unknown.  Armed with
knowledge of the Fermi liquid parameters, we would have a complete
parameterization of the low-energy excitations of the fully interacting
electron gas.  This would in turn allow nearly all thermodynamic, response,
and transport properties to be determined quantitatively,\cite{giuliani}
enabling us to understand the precise role that correlation plays in the
behavior of the HEG\@.

In this work we use quantum Monte Carlo (QMC)
calculations\cite{ceperley_1980,foulkes_2001} to determine the Fermi liquid
parameters of the two-dimensional (2D) HEG\@. Specifically, we have employed
the variational Monte Carlo (VMC) and diffusion Monte Carlo (DMC)
methods.\cite{foulkes_2001} VMC calculations involve taking the expectation
value of a many-electron Hamiltonian with respect to a trial wave function
that can be of arbitrary complexity. In our work, the trial wave function was
optimized by minimizing first the variance of the
energy,\cite{umrigar_1988a,ndd_newopt} then the energy expectation
value\cite{umrigar_emin} with respect to wave-function parameters. In
DMC\cite{ceperley_1980} we simulate a process governed by the Schr\"{o}dinger
equation in imaginary time in order to project out the ground-state component
of an initial wave function.  We use the fixed-node
approximation\cite{anderson_1976} to impose fermionic antisymmetry.  All our
QMC calculations were performed using the \textsc{casino}
code.\cite{casino_ref}

In Refs.\ \onlinecite{ndd_band} and \onlinecite{ndd_effmass2} we presented DMC
calculations of the 2D HEG single-particle energy band ${\cal E}(k)$, enabling
us to predict the quasiparticle effective mass $m^\ast$.  In the present work
we use DMC calculations to determine the Landau interaction
functions\cite{giuliani} and hence Fermi liquid parameters.  Our approach is
similar to that of the pioneering work of Kwon \textit{et
  al.},\cite{kwon_1994} which was undertaken eighteen years ago and is, to our
knowledge, the only previous attempt to calculate the Fermi liquid parameters
directly using QMC\@.  Kwon \textit{et al.}\ were unable to obtain consistent
quantitative results, primarily because of the extremely small system sizes
that they were forced to use at that time.  However, there have been enormous
developments in QMC methodology and computer power in the last two decades,
and the time has come to revisit this grand-challenge problem.  The major
causes of computational expense in this work are (i) the need to overcome the
finite-size errors in the Fermi liquid parameters by performing calculations
at a range of system sizes and (ii) the fact that even for small numbers of
electrons it is necessary to take the difference of very similar total
energies to obtain the interaction functions.  Point (ii) makes every aspect
of this work computationally expensive: not only does each QMC calculation
have to be sufficiently long that the statistical error bars are small
compared with the differences to be resolved, but it must be ensured that the
trial wave function is very highly optimized.  These calculations were only
made possible by access to the Jaguar machine at Oak Ridge Leadership
Computing Facility.

The rest of this paper is structured as follows.  In
Sec.\ \ref{sec:landau_e_fnal} we give an overview of the relevant aspects of
Fermi liquid theory and describe our computational approach to the problem.
Our results are presented in Sec.\ \ref{sec:results}.  Finally, we draw our
conclusions in Sec.\ \ref{sec:conclusions}. We use Hartree atomic units, in
which the Dirac constant, the electronic charge and mass, and $4\pi$ times the
permittivity of free space are unity ($\hbar=|e|=m_e=4\pi\epsilon_0=1$),
throughout.

\section{Evaluating the Landau energy functional \label{sec:landau_e_fnal}}

\subsection{Parameterization of excitation energies}

The Landau energy functional\cite{giuliani} is a parameterization of the
energies of the ground state and low-lying excited states of the HEG:
\begin{eqnarray} E & = & E_0+\sum_{{\bf k},\sigma} {\cal E}_\sigma({\bf k})
  \delta {\cal N}_{{\bf k},\sigma} \nonumber \\ & & {} + \frac{1}{2}
  \sum_{({\bf k},\sigma) \neq ({\bf k}^\prime,\sigma^\prime)}
  f_{\sigma,\sigma^\prime}({\bf k},{\bf k}^\prime) \delta {\cal N}_{{\bf
      k},\sigma} \delta {\cal N}_{{\bf
      k}^\prime,\sigma^\prime}, \label{eq:landau_e_fnal}
  \end{eqnarray}
where $\delta {\cal N}_{{\bf k},\sigma}$ is the change to the ground-state
quasiparticle occupation number for wavevector ${\bf k}$ and spin $\sigma$,
and $E_0$ is the ground-state energy.  Sufficiently close to the Fermi
surface, the energy band ${\cal E}_\sigma({\bf k})$ is linear in $k$ and hence
we may write
\begin{equation} {\cal E}_\sigma({\bf k}) = {\cal
  E}_F+\frac{k_F}{m^\ast}(k-k_F), \label{eq:qem_defn} \end{equation} where
${\cal E}_F$ is the Fermi energy, $k_F$ is the Fermi wavevector, and $m^\ast$
is the quasiparticle effective mass.  The Landau interaction function
$f_{\sigma,\sigma^\prime}({\bf k},{\bf k}^\prime)$ describes energy
contributions arising from pairs of quasiparticles.  Close to the Fermi
surface we may neglect the dependence of $f$ on the magnitudes of the
wavevectors and write the Landau interaction functions as $f_{\sigma
  \sigma^\prime}(\theta_{{\bf k}{\bf k}^\prime})$, where $\theta_{{\bf k}{\bf
    k}^\prime}$ is the angle between ${\bf k}$ and ${\bf k}^\prime$.  The
$l$th Fermi liquid parameter of the paramagnetic HEG is defined
as\cite{giuliani}
\begin{equation} F_l^{s,a}=\frac{AN^\ast_p(0)}{4\pi}\int_0^{2\pi} \left[ f_{\uparrow
      \uparrow}(\theta_{{\bf kk}^\prime}) \pm f_{\uparrow
      \downarrow}(\theta_{{\bf kk}^\prime}) \right] \cos(l\theta) \,
  d\theta, \label{eq:FLP_para} \end{equation} where $A=\pi r_s^2N$ is the area
of the simulation cell, $r_s$ is the radius of the circle that contains one
electron on average, $N$ is the number of electrons in the simulation cell,
and $N^\ast_p(0)=m^\ast/\pi$ is the quasiparticle density of states per unit
area at the Fermi surface.  The suffixes $s$ and $a$ (for ``symmetric'' and
``antisymmetric'') correspond to addition and subtraction in the integrand,
respectively.  For a fully ferromagnetic HEG, the $l$th Fermi liquid parameter
is defined as
\begin{equation} F_l=\frac{AN^\ast_f(0)}{2\pi}\int_0^{2\pi} f_{\uparrow
      \uparrow}(\theta_{{\bf kk}^\prime}) \cos(l\theta) \,
  d\theta, \label{eq:FLP_ferro} \end{equation} where the quasiparticle density
of states per unit area is $N^\ast_f(0)=m^\ast/(2\pi)$.

\subsection{Hartree-Fock theory \label{sec:HF}}

The total energy of a finite HEG in Hartree-Fock theory can be written as
\begin{equation} E_{\rm HF} = \sum_\sigma \sum_{{\bf k} \in {\rm Occ}_\sigma}
  \frac{k^2}{2} - \frac{1}{2} \sum_\sigma \sum_{{\bf k} \neq {\bf k}^\prime
    \in {\rm Occ}_\sigma} \frac{2 \pi}{A |{\bf k}-{\bf k}^\prime|} + \frac{N
    v_M}{2}, \label{eq:HF_energy_finite} \end{equation} where $v_M$ is the
Madelung constant, $N$ is the number of electrons, $A$ is the area of the
simulation cell and ${\rm Occ}_\sigma$ is the set of occupied states for spin
$\sigma$.  The Hartree-Fock energy is already in the form of the Landau energy
functional and hence within Hartree-Fock theory the energy band is
\begin{equation} {\cal E}_\sigma({\bf k})=\frac{k^2}{2}-\sum_{{\bf k}^\prime
    \in {\rm GS}_\sigma} \frac{2 \pi}{A |{\bf k}-{\bf
      k}^\prime|} \end{equation} for ground-state unoccupied wavevectors,
where ${\rm GS}_\sigma$ is the set of states occupied in the ground state, and
\begin{equation} {\cal E}_\sigma({\bf k})=\frac{k^2}{2}-\sum_{{\bf k}^\prime \neq
    {\bf k} \in {\rm GS}_\sigma} \frac{2 \pi}{A |{\bf k}-{\bf
      k}^\prime|} \end{equation} for ground-state occupied wavevectors.  It
also follows immediately from Eq.\ (\ref{eq:HF_energy_finite}) that the Landau
interaction functions in Hartree-Fock theory are
\begin{equation} f_{\sigma \sigma}({\bf k},{\bf k}^\prime)=
  -\frac{2 \pi}{A |{\bf k}-{\bf k}^\prime|}
  \delta_{\sigma,\sigma^\prime}. \end{equation}

For excitations close to the Fermi surface, $|{\bf k}| \approx |{\bf
  k}^\prime|\approx k_F$, where $k_F$ is the Fermi wavevector.  Let
$\theta_{{\bf k}{\bf k}^\prime}$ be the angle between ${\bf k}$ and ${\bf
  k}^\prime$.  Then $|{\bf k}-{\bf k}^\prime|^2=2k_F^2[1-\cos(\theta_{{\bf
      k}{\bf k}^\prime})]$.  $A=\pi r_s^2N$ and, for a paramagnetic HEG,
$k_F=\sqrt{2}/r_s$, so
\begin{equation} Nf_{\sigma \sigma^\prime}({\bf k},{\bf k}^\prime)=\frac{-
    \delta_{\sigma \sigma^\prime}}{r_s\sqrt{1-\cos(\theta_{{\bf k}{\bf
          k}^\prime})}}. \label{eq:hf_landau_para} \end{equation} For a
ferromagnetic HEG, $k_F=2/r_s$ and so
\begin{equation} Nf_{\uparrow \uparrow}({\bf k},{\bf
    k}^\prime)=\frac{-1}{r_s\sqrt{2-2\cos(\theta_{{\bf k}{\bf
          k}^\prime})}}. \label{eq:hf_landau_ferro} \end{equation}

\subsection{QMC calculations \label{sec:qmc_calcs}}

\subsubsection{Evaluating the Landau interaction functions}

By Eq.\ (\ref{eq:landau_e_fnal}) we can evaluate the Landau interaction
functions at a discrete set of angles $\{\theta_i\}$ as
\begin{equation} f_{\sigma
  \sigma^\prime}(\theta_i)=-[E_{\sigma \sigma^\prime}({\bf k}_i,{\bf
      k}_i^\prime)+E_0-E_+({\bf k}_i)-E_-({\bf
      k}_i^\prime)], \label{eq:intfn_eval} \end{equation} where $E_0$ is the
ground-state energy, $E_{\sigma \sigma^\prime}({\bf k}_i,{\bf k}_i^\prime)$ is
the total energy of an excited state in which an electron is promoted from
${\bf k}_i^\prime$ near the Fermi surface to ${\bf k}_i$ just above the Fermi
surface, $\theta_i$ is the angle between ${\bf k}_i$ and ${\bf k}_i^\prime$,
and $\sigma$ and $\sigma^\prime$ are the corresponding spins.  $E_+({\bf
  k}_i)$ and $E_-({\bf k}_i^\prime)$ are the total energies of the system with
an electron added to ${\bf k}_i$ and removed from ${\bf k}_i^\prime$,
respectively.

\subsubsection{Finite-size errors}

Our QMC calculations were performed for electron gases in finite cells subject
to periodic boundary conditions.  The available wavevectors $\{{\bf k}\}$ are
therefore the reciprocal lattice points of the simulation cell.  The use of a
finite cell prevents the description of long-range Coulomb and correlation
effects,\cite{holzmann_2009,holzmann_2011} giving rise to finite-size errors
in the Fermi liquid parameters.  We have calculated the Landau interaction
functions via total-energy differences in finite cells, used these to evaluate
the Fermi liquid parameters, then extrapolated the parameters to the
thermodynamic limit, where they should become independent of the choice of
simulation cell and the precise excitations made to determine the interaction
functions.

\subsubsection{Simulation cell}

In our calculations we have used square simulation cells with simulation-cell
Bloch vector\cite{rajagopal_1994,rajagopal_1995} ${\bf k}_s={\bf 0}$.  There
exist quantities such as the ground-state total energy, pair-correlation
function, and static structure factor that can be twist
averaged\cite{lin_twist_av} in the conventional sense (i.e., one can evaluate
estimators for these quantities at different simulation-cell Bloch vectors
${\bf k}_s$ and then average the results).  However, there are other
quantities such as the momentum density, the energy band, and the Landau
interaction functions for which ${\bf k}_s$ determines the set of wavevectors
at which the quantities are defined in a finite cell, so that by using
different ${\bf k}_s$ we may obtain additional points on the quantity as a
function of wavevector.  The computational effort required to obtain
excitation energies at different ${\bf k}_s$ is essentially the same as the
computational effort required to obtain excitation energies by considering
completely different excitations.  Since the latter approach provides data
that are in some sense more independent, we concluded that our computational
effort was better invested in studying different excitations as opposed to
changing ${\bf k}_s$.

The number of electrons in the ground state was chosen to be a ``magic
number,'' corresponding to a closed-shell configuration in each case.  For
ferromagnetic HEGs, our calculations were performed with $N=29$, $57$, and
$101$ electrons in the ground state.  For paramagnetic HEGs our calculations
were performed with $N=26$, $50$, $74$, and $114$ electrons in the ground
state.

The simulation-cell area was held constant when electrons were added to or
removed from the ground-state configuration.  In the case of noninteracting
electrons this gives the energy band ($k^2/2$) and Landau interaction
functions (zero) exactly without finite-size error.  (Note that in the
free-electron model there are finite-size errors due to momentum quantization
in the total energy, but no finite-size errors in the excitation energies.)
For interacting electrons, the fact that the density changes when electrons
are added or subtracted may be a source of finite-size error, but the error
involved is certainly much smaller than the error that would result from
allowing the cell area to change.

\subsubsection{Trial wave function \label{sec:trial_wf}}

We used trial wave functions of Slater-Jastrow-backflow
form.\cite{ndd_jastrow,kwon_bf,backflow} More detailed information about our
trial wave functions can be found in Ref.\ \onlinecite{ndd_effmass2}.  In
Ref.\ \onlinecite{ndd_band} we argued that our DMC calculations for the 2D HEG
retrieve more than 99\% of the correlation energy.

The DMC time steps used in our calculations were $0.04$, $0.2$, and $0.4$
a.u.\ at $r_s=1$, 5, and 10, respectively, for paramagnetic HEGs, and $0.01$,
$0.2$, and $0.4$ a.u.\ at $r_s=1$, 5, and 10, respectively, for ferromagnetic
HEGs.

At $r_s=5$ we find the VMC energy variance per electron to be $1.48 \times
10^{-4}$ and $2.44 \times 10^{-5}$ a.u.\ for paramagnetic and ferromagnetic
HEGs, respectively.  Thus our trial wave functions are considerably more
accurate for ferromagnetic HEGs, in which exchange effects are dominant.  This
suggests that it might be advantageous to use pairing (geminal) orbitals for
opposite-spin electrons in paramagnetic HEGs.\cite{geminal} Another
possibility for improving the trial wave function would be to use different
Jastrow factors\cite{bouabca} and backflow functions for each shell of
plane-wave orbitals.  However, given the expense of our calculations, there is
at present little scope for using more sophisticated wave-function forms.

\section{Results \label{sec:results}}

\subsection{Landau interaction functions \label{sec:int_fns}}

\begin{figure}
\begin{center}
\includegraphics[clip,width=0.4\textwidth]{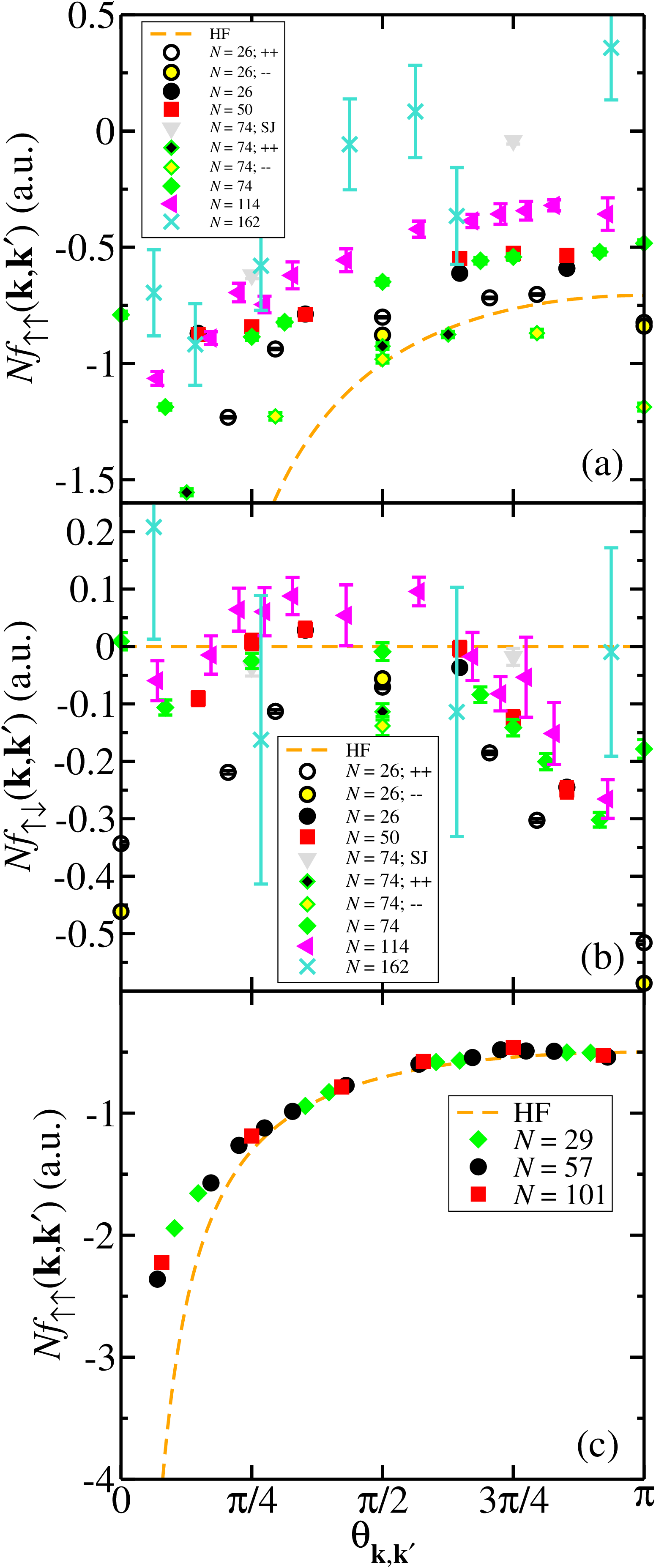}
\caption{(Color online) Landau interaction functions $f_{\sigma
    \sigma^\prime}({\bf k},{\bf k}^\prime)$ for 2D HEGs of density parameter
  $r_s=1$, multiplied by system size $N$.  The parallel- and antiparallel-spin
  interaction functions in a paramagnetic HEG are shown in panels (a) and (b),
  respectively, while the interaction function in a fully ferromagnetic HEG is
  shown in panel (c).  All results were obtained with a
  Slater-Jastrow-backflow trial wave function, except where labeled ``SJ,'' in
  which case a Slater-Jastrow wave function was used. The results labeled
  ``++'' were obtained in double-addition calculations, whereas those labeled
  ``$--$'' were obtained in double-subtraction calculations.  All other
  results were obtained by promoting a single electron, leaving a hole.  Note
  that the data points within each curve are correlated: for example, the
  interaction-function values at a given system size all depend on the DMC
  estimate of the ground-state energy. For comparison, we show the
  Hartree-Fock (HF) ``interaction functions'' for infinite system size
  [Eqs.\ (\ref{eq:hf_landau_para}) and
    (\ref{eq:hf_landau_ferro})]. \label{fig:int_params_rs1}}
\end{center}
\end{figure}

\begin{figure}
\begin{center}
\includegraphics[clip,width=0.4\textwidth]{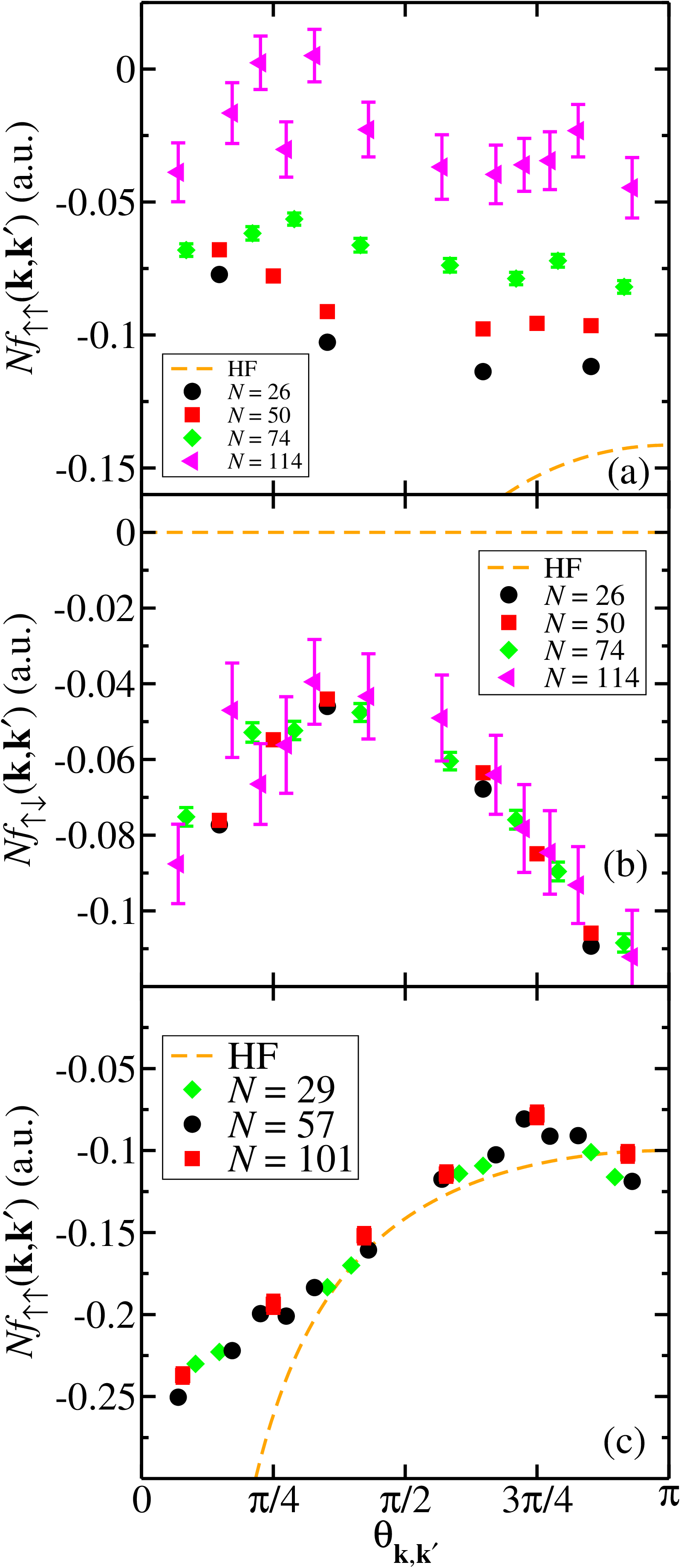}
\caption{(Color online) As Fig.\ \ref{fig:int_params_rs1}, but for HEGs at
  $r_s=5$. \label{fig:int_params_rs5}}
\end{center}
\end{figure}

\begin{figure}
\begin{center}
\includegraphics[clip,width=0.4\textwidth]{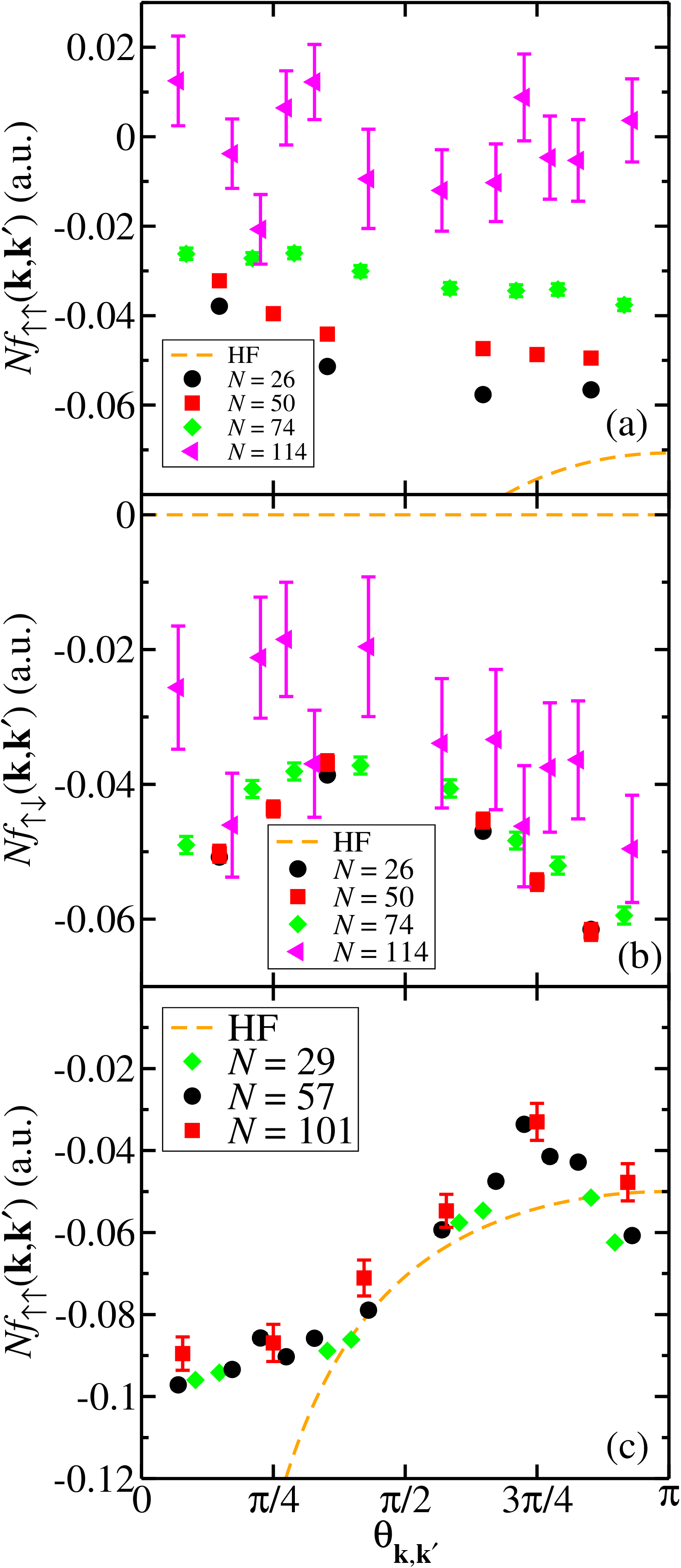}
\caption{(Color online) As Fig.\ \ref{fig:int_params_rs1}, but for HEGs at
  $r_s=10$. \label{fig:int_params_rs10}}
\end{center}
\end{figure}

The DMC Landau interaction functions at $r_s=1$, $5$, and $10$ are shown in
Figs.\ \ref{fig:int_params_rs1}, \ref{fig:int_params_rs5}, and
\ref{fig:int_params_rs10}, respectively.  The statistical error bars are very
much smaller for ferromagnetic HEGs than for paramagnetic HEGs, reflecting the
relative accuracy of the trial wave functions in the two cases (see
Sec.\ \ref{sec:trial_wf}).  Note that the data points are correlated, and so
the error bars should be interpreted with caution.  The Hartree-Fock
interaction function (i.e., the exchange interaction: see Sec.\ \ref{sec:HF})
is reasonably accurate at large $\theta$, but the parallel-spin interaction
function is pathological as $\theta\rightarrow 0$ due to the lack of
screening.

The differences between the Slater-Jastrow and Slater-Jastrow-backflow DMC
data are significant, confirming that the former are insufficiently accurate,
as argued in Ref.\ \onlinecite{kwon_1994}.  There is a significant difference
between the Landau interaction functions obtained from DMC calculations in
which a single electron is promoted and calculations in which pairs of
electrons are added or pairs of electrons are removed from the ground-state
configuration.  Finite-size effects in promotion energies are expected to be
smaller because the density of the HEG in a finite cell is unchanged by such
excitations, unlike double additions or subtractions.  We have used only
promotions in our production calculations to determine the Fermi liquid
parameters.

\subsection{Fermi liquid parameters \label{sec:evaluating_flps}}

\subsubsection{Numerical integration of the Landau interaction functions to
  find the Fermi liquid parameters}

By Eq.\ (\ref{eq:FLP_para}), the Fermi liquid parameters divided by the
effective mass $m^\ast$ for an $N$-electron paramagnetic HEG of density
parameter $r_s$ can be written as
\begin{eqnarray} \frac{F_l^{s,a}}{m^\ast} & = & \frac{r_s^2 N}{2\pi}
  \int_0^\pi \left[ f_{\uparrow \uparrow}(\theta) \pm f_{\uparrow
      \downarrow}(\theta) \right] \cos(l\theta) \, d\theta \nonumber \\ & = &
  \frac{r_s^2}{4} \left[ a_l^{\uparrow \uparrow} \pm a_l^{\uparrow \downarrow}
    \right], \label{eq:FLP_from_FT} \end{eqnarray} where
\begin{equation} a_l^{\sigma \sigma^\prime} = \frac{2}{\pi} \int_0^{\pi}
  Nf_{\sigma \sigma^\prime}(\theta) \cos(l\theta) \,
  d\theta \label{eq:ft_intfn} \end{equation} is the $l$th Fourier component of
$Nf_{\sigma \sigma^\prime}(\theta)$.  To evaluate these Fourier components we
use Simpson's rule (integration of piecewise quadratic interpolants)
generalized for the case of a nonuniform integration grid.  The set of angles
$\{\theta_i\}$ at which the integrand is available does not generally include
the endpoints of the integration region ($0$ and $\pi$).  Where necessary we
integrate a straight-line interpolation of the closest two data points up to
the endpoints of the integration region.

The Fourier coefficients $\{a_l^{\sigma \sigma^\prime}\}$ are linear in $E_0$,
$E_+({\bf k}_i)$, $E_-({\bf k}_i^\prime)$, and $E_{\sigma \sigma^\prime}({\bf
  k}_i,{\bf k}_i^\prime)$.  We therefore gather the coefficients of each of
these in the expression for $a_l^{\sigma \sigma^\prime}$
[Eqs.\ (\ref{eq:ft_intfn}) and (\ref{eq:intfn_eval})].  Finally, we evaluate
the coefficients of $E_0$, $E_+({\bf k}_i)$, $E_-({\bf k}_i^\prime)$, and
$E_{\sigma \sigma^\prime}({\bf k}_i,{\bf k}_i^\prime)$ in the expression for
$F_l^{s,a}/m^\ast$ using Eq.\ (\ref{eq:FLP_from_FT}).  Since we have
independent DMC estimates of $E_0$, $E_+({\bf k}_i)$, $E_-({\bf k}_i^\prime)$,
and $E_{\sigma \sigma^\prime}({\bf k}_i,{\bf k}_i^\prime)$, we can evaluate
both the expected Fermi liquid parameters divided by effective mass and the
accompanying standard errors.

A systematic integration error arises from the use of numerical quadrature
with a finite set of angles.  This integration error is not included in our
statistical error bars.  We may place an upper bound on the error by comparing
the Fermi liquid parameters obtained using (i) the generalized composite
Simpson's rule and (ii) the generalized composite trapezoidal rule to obtain
$a_l^{\sigma \sigma^\prime}$.  Results are given in Table
\ref{table:comp_int_rules_rs5} for HEGs at $r_s=5$.  It is clear that the
integration error is negligible compared with the random error for
$F_0/m^\ast$ and $F_1/m^\ast$.  For $F_2^{s,a}/m^\ast$ the effect of the
choice of integration rule is more significant, especially for smaller numbers
of electrons, where relatively few values of $\theta$ are available; however
the error is still small compared with the overall results.

\begin{table}
\caption{Comparison of the first three Fermi liquid parameters (FLPs) in
  a.u.\ for a 74-electron paramagnetic HEG at $r_s=5$ ($F_0^s$, $F_0^a$,
  $F_1^s$, $F_1^a$, $F_2^s$, and $F_2^a$) and a 57-electron ferromagnetic HEG
  at $r_s=5$ ($F_0$, $F_1$, and $F_2$), derived from the Fourier components of
  the Slater-Jastrow-backflow DMC Landau interaction functions by numerical
  integration using the composite Simpson's rule and the composite trapezoidal
  rule. \label{table:comp_int_rules_rs5}}
\begin{tabular}{cr@{.}lr@{.}l}
\hline \hline

FLP over eff.\ mass & \multicolumn{2}{c}{Simpson} &
\multicolumn{2}{c}{Trapezoidal} \\

\hline

$F_0^s/m^\ast$ & $-1$&$78(4)$ & ~~$-1$&$78(4)$ \\

$F_0^a/m^\ast$ & $0$&$01(1)$  & $0$&$01(1)$  \\

$F_1^s/m^\ast$ & $0$&$18(1)$  & $0$&$18(1)$  \\

$F_1^a/m^\ast$ & $-0$&$06(1)$ & $-0$&$06(1)$ \\

$F_2^s/m^\ast$ & $-0$&$18(1)$ & $-0$&$22(1)$ \\

$F_2^a/m^\ast$ & $0$&$12(1)$  & $0$&$12(1)$  \\

$F_0/m^\ast$ & $-1$&$94(1)$  & ~~$-1$&$94(1)$ \\

$F_1/m^\ast$ & $-0$&$461(2)$ & $-0$&$466(3)$ \\

$F_2/m^\ast$ & $-0$&$170(3)$ & $-0$&$193(3)$ \\

\hline \hline
\end{tabular}
\end{table}

The Fermi liquid parameters divided by the effective mass for a ferromagnetic
HEG can be written as
\begin{equation} \frac{F_l}{m^\ast}=\frac{r_s^2 N}{2\pi} \int_0^\pi
    f_{\uparrow \uparrow}(\theta) \cos(l\theta) \, d\theta = \frac{r_s^2}{4}
    a_l^{\uparrow \uparrow}. \label{eq:FLP_from_FT_ferro} \end{equation} Hence
we can evaluate the Fermi liquid parameters divided by the effective mass
(with standard errors) for the ferromagnetic case using the same approach as
for the paramagnetic case.

\subsubsection{Finite-size extrapolation of the Fermi liquid parameters}

\begin{figure}
\begin{center}
\includegraphics[clip,width=0.4\textwidth]{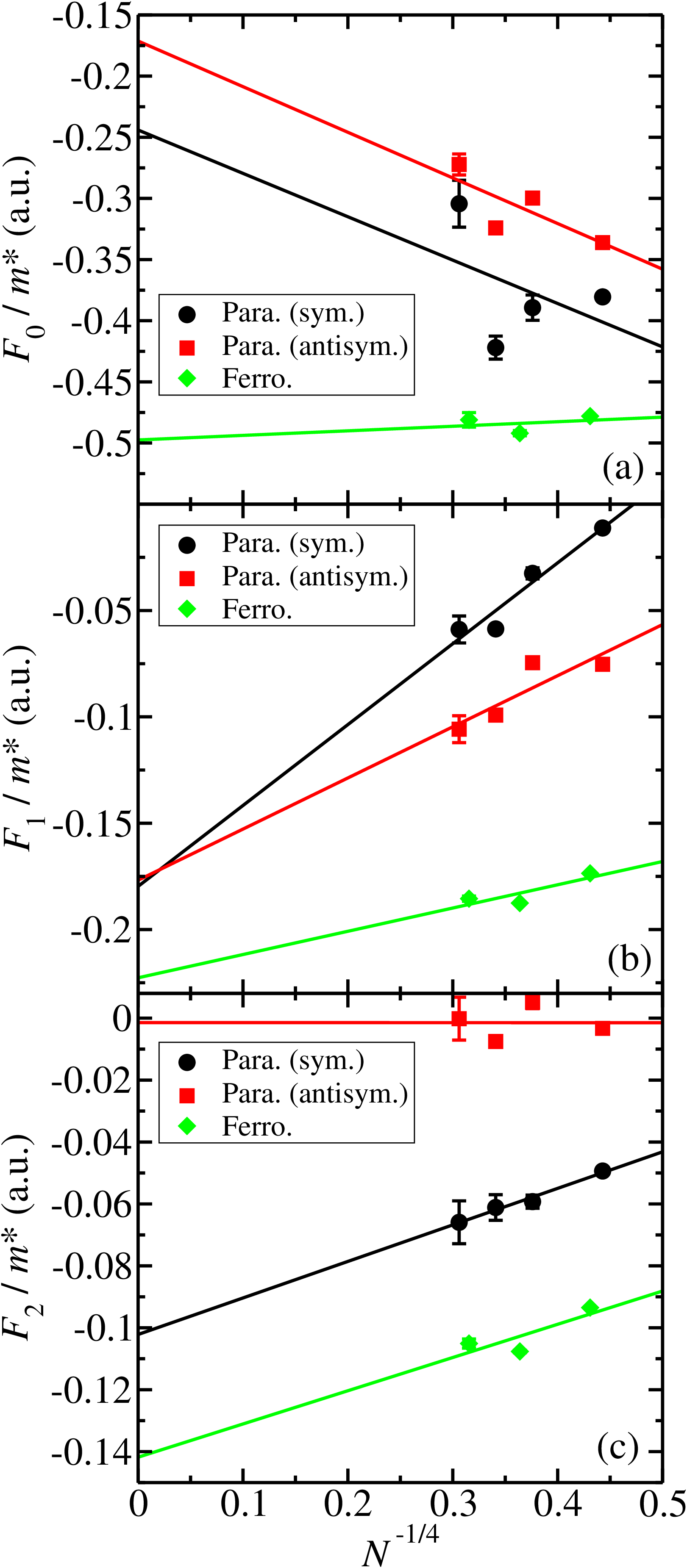}
\caption{(Color online) Fermi liquid parameters divided by effective mass, (a)
  $F_0/m^\ast$, (b) $F_1/m^\ast$, and (c) $F_2/m^\ast$, against system size
  for paramagnetic HEGs (both symmetric and antisymmetric parameters) and
  ferromagnetic HEGs of density parameter $r_s=1$.
\label{fig:F_extrap_ert_N_rs1}}
\end{center}
\end{figure}

\begin{figure}
\begin{center}
\includegraphics[clip,width=0.4\textwidth]{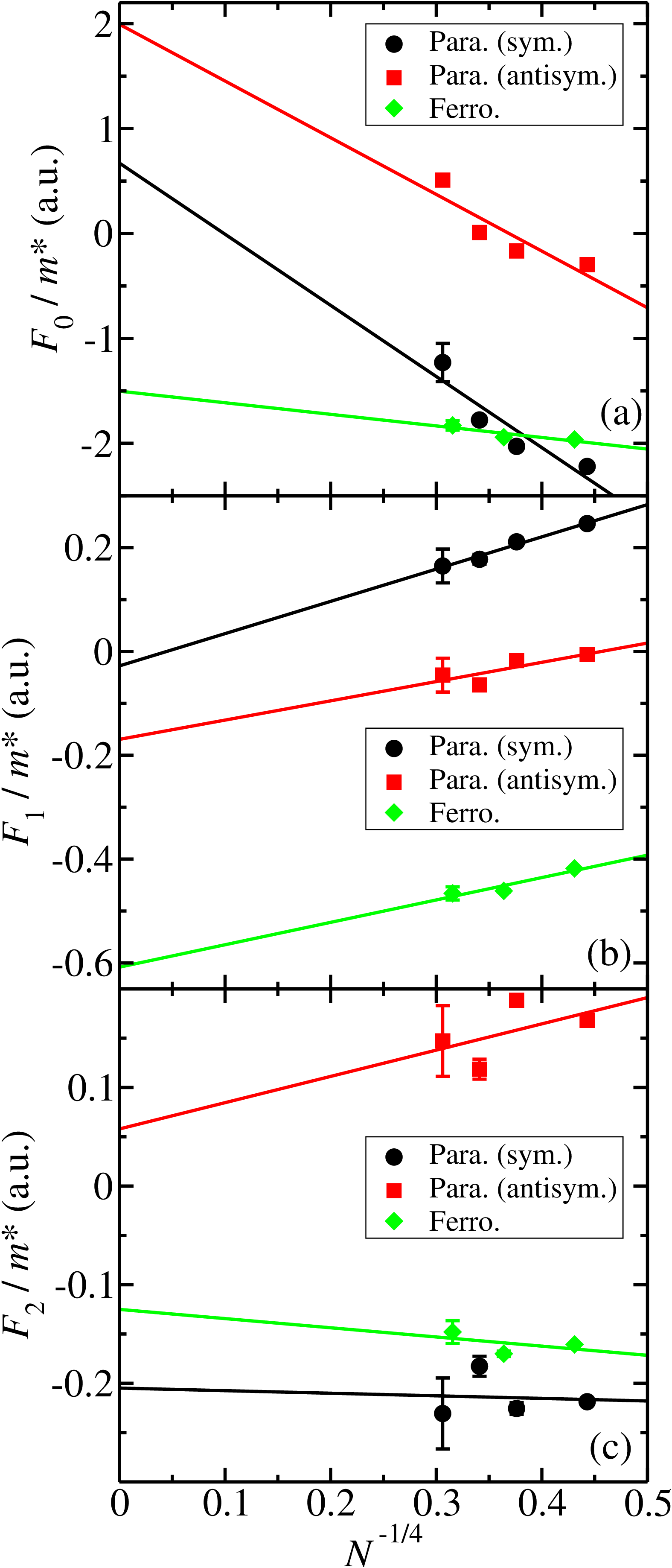}
\caption{(Color online) As Fig.\ \ref{fig:F_extrap_ert_N_rs1} but for HEGs of
  density parameter $r_s=5$. \label{fig:F_extrap_ert_N_rs5}}
\end{center}
\end{figure}

\begin{figure}
\begin{center}
\includegraphics[clip,width=0.4\textwidth]{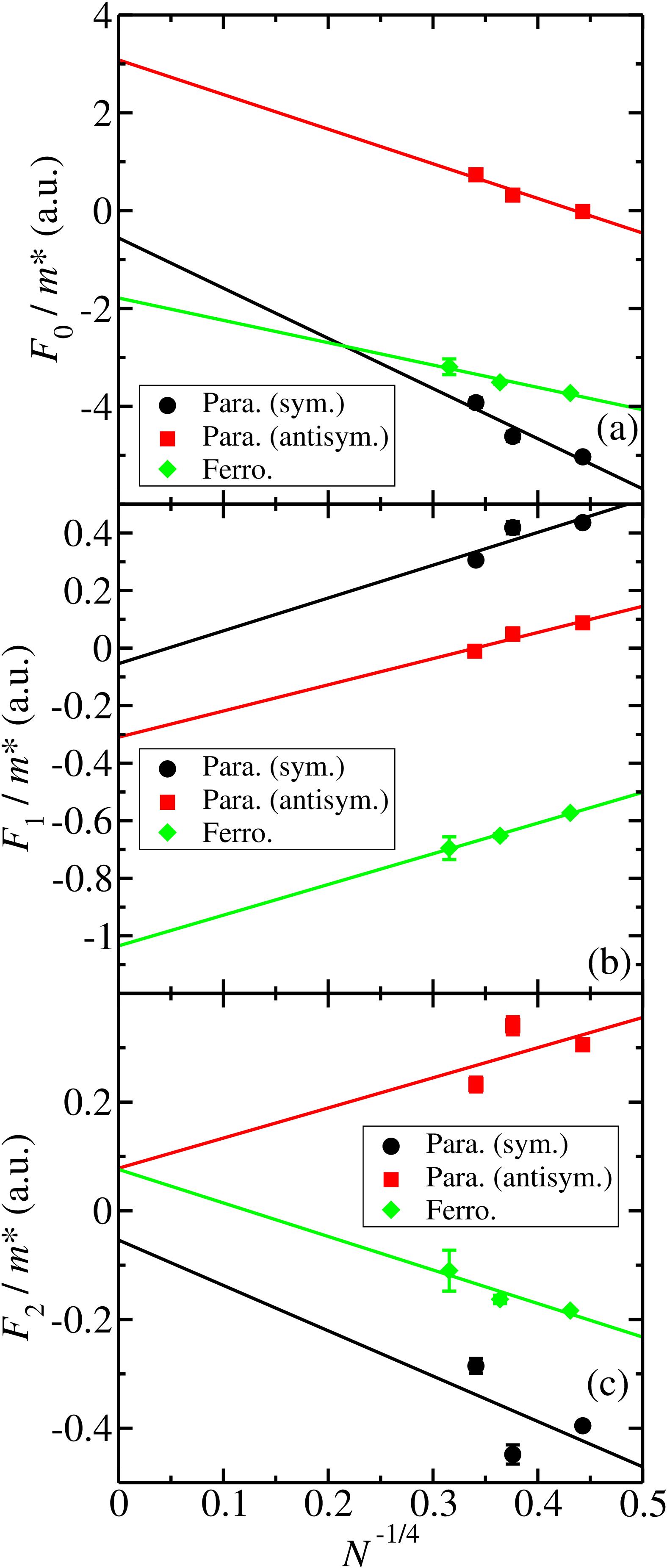}
\caption{(Color online) As Fig.\ \ref{fig:F_extrap_ert_N_rs1} but for HEGs of
  density parameter $r_s=10$. \label{fig:F_extrap_ert_N_rs10}}
\end{center}
\end{figure}

We have calculated the Fermi liquid parameters divided by the effective mass
at three or more system sizes for each density and magnetic state and we have
extrapolated the parameters to infinite system size by assuming the
finite-size error falls off as $N^{-1/4}$.  This was the scaling found for
Fermi liquid properties in Refs.\ \onlinecite{holzmann_2009} and
\onlinecite{holzmann_2011}.  The finite-size errors in the Fermi liquid
parameters are determined by long-range correlation effects and we have
therefore used the same system-size scaling for paramagnetic and ferromagnetic
phases.  The finite-size extrapolations at $r_s=1$, $5$, and $10$ are shown in
Figs.\ \ref{fig:F_extrap_ert_N_rs1}, \ref{fig:F_extrap_ert_N_rs5}, and
\ref{fig:F_extrap_ert_N_rs10}, respectively.  The statistical error bars on
the Fermi liquid parameters divided by the effective mass are generally much
smaller than the apparent fluctuations as a function of system size.  These
fluctuations are presumably finite-size effects arising from the discrete
nature of the lattice of wavevectors.  We have therefore decided not to weight
the residuals by the inverse of the error bars when performing the
extrapolation to infinite system size.  The quoted error bars on the
extrapolated Fermi liquid parameters divided by the effective mass given in
Tables \ref{table:fermi_liq_params_para} and
\ref{table:fermi_liq_params_ferro} are obtained from an ordinary least-squares
fit.

There is no evidence for a systematic deviation of the Fermi liquid parameters
from the fitted curves at small system sizes beyond the ``noise'' that
obviously affects all the data points shown in
Figs.\ \ref{fig:F_extrap_ert_N_rs1}, \ref{fig:F_extrap_ert_N_rs5}, and
\ref{fig:F_extrap_ert_N_rs10}.  We have therefore included all the data shown
in these figures in our extrapolation to infinite system size.

We have attempted to check the exponent used for finite-size extrapolation of
our Fermi liquid parameters by simultaneously fitting the functions
\begin{equation} F_i(N)=c_i+a_iN^\gamma \end{equation}
to all our DMC data using a $\chi^2$ fit.  The values of the parameters $c_i$
and $a_i$ differed for each Fermi liquid parameter at each density, while the
exponent $\gamma$ was constrained to be the same in all cases.  We find the
optimal exponent to be $\gamma=-0.24(10)$, which is superficially consistent
with the exponent $\gamma=-0.25$ determined theoretically by Holzmann
\textit{et al.}\cite{holzmann_2009} However, this is certainly not a
conclusive numerical determination of the exponent $\gamma$.  The $\chi^2$
value per data point with the optimal exponent of $\gamma=-0.24$ is $6.88$.
The fact that this is much greater than 1 indicates that the fit is far from
perfect.  The main reason is the finite-size ``noise'' due to shell-filling
effects, which is not included in the statistical error bars on the Fermi
liquid parameters over effective mass.  (The error bars only account for the
random noise inherent in the QMC calculation.)  If, instead of performing a
$\chi^2$ fit, one performs a simple least-squares fit (i.e., each data point
is weighted equally rather than by the squared reciprocal of the nominal error
bar), one finds that the optimal exponent is $\gamma=-1(2)$.  Alternatively,
if one performs a $\chi^2$ fit with the exponent fixed at $\gamma=-1$, the
resulting $\chi^2$ value is $7.26$ per data point, which is only slightly
larger than the $\chi^2$ obtained with the ``optimal'' value of
$\gamma=-0.24$.

In summary, we do not believe that we can meaningfully determine the
finite-size scaling exponent $\gamma$ numerically, but our results are
consistent with the value $\gamma=-0.25$ determined theoretically by Holzmann
\textit{et al.}  We have therefore used this value in our analysis.

\subsubsection{DMC results for the Fermi liquid parameters}

Our results for the first three Fermi liquid parameters (symmetric and
antisymmetric) of the paramagnetic HEG divided by the effective mass are given
in Table \ref{table:fermi_liq_params_para}, and the analogous results for a
ferromagnetic HEG are given in Table \ref{table:fermi_liq_params_ferro}.

\begin{table*}
\begin{center}
\caption{Fermi liquid parameters over effective mass in a.u.\ for the
  paramagnetic 2D HEG, extrapolated to the thermodynamic
  limit. \label{table:fermi_liq_params_para}}
\begin{tabular}{cr@{}lr@{}lr@{}lr@{}lr@{}lr@{}l} \\ \hline \hline

$r_s$ & \multicolumn{2}{c}{$F_0^s/m^\ast$} &
  \multicolumn{2}{c}{$F_0^a/m^\ast$} & \multicolumn{2}{c}{$F_1^s/m^\ast$} &
  \multicolumn{2}{c}{$F_1^a/m^\ast$} & \multicolumn{2}{c}{$F_2^s/m^\ast$} &
  \multicolumn{2}{c}{$F_2^a/m^\ast$} \\

\hline

$1$ & $-0$&$.2(2)$ & $-0$&$.17(8)$ & $-0$&$.18(3)$ & $-0$&$.18(4)$ &
$-0$&$.102(5)$ & $0$&$.00(2)$ \\

$5$ & $0$&$.7(8)$ & $2$&$.0(7)$ & $-0$&$.03(2)$ & $-0$&$.17(7)$ & $-0$&$.2(1)$
& $0$&$.1(1)$ \\

$10$ & $-1$&$(1)$ & $3$&$.1(7)$ & $-0$&$.1(3)$ & $-0$&$.3(1)$ & $-0$&$.1(5)$ &
$0$&$.1(4)$ \\

\hline\hline
\end{tabular}
\end{center}
\end{table*}

\begin{table}
\begin{center}
\caption{Fermi liquid parameters over effective mass in a.u.\ for the
  ferromagnetic 2D HEG, extrapolated to the thermodynamic
  limit. \label{table:fermi_liq_params_ferro}}
\begin{tabular}{cr@{}lr@{}lr@{}l} \\ \hline \hline

$r_s$ & \multicolumn{2}{c}{$F_0/m^\ast$} & \multicolumn{2}{c}{$F_1/m^\ast$} &
  \multicolumn{2}{c}{$F_2/m^\ast$} \\

\hline

$1$ & $-0$&$.50(4)$ & $-0$&$.22(3)$ & $-0$&$.14(3)$ \\

$5$ & $-1$&$.5(2)$ & $-0$&$.61(6)$ & $-0$&$.13(6)$ \\

$10$ & $-1$&$.8(3)$ & $-1$&$.03(3)$ & $0$&$.08(8)$ \\

\hline\hline
\end{tabular}
\end{center}
\end{table}

\subsection{Relationships between the Fermi liquid parameters and other
  accessible quantities}

\subsubsection{Quasiparticle effective mass}

By Eq.\ (\ref{eq:qem_defn}), the quasiparticle effective mass is
\begin{equation}  m^\ast=\frac{k_F}{(d{\cal
      E}/dk)_{k_F}}. \label{eq:qem_eval} \end{equation} However, by Galilean
invariance,\cite{giuliani} the Fermi liquid parameter $F_1^s$ is related to
the effective mass of the paramagnetic HEG via $m^\ast=1+F_1^s$, and, for the
ferromagnetic HEG, $m^\ast=1+F_1$.  Hence
\begin{equation}
  m^\ast=\frac{1}{1-F_1^s/m^\ast}, \label{eq:qem_flp_eval_para} \end{equation}
for paramagnetic HEGs and
\begin{equation}
  m^\ast=\frac{1}{1-F_1/m^\ast}, \label{eq:qem_flp_eval_ferro} \end{equation}
for ferromagnetic HEGs, so we can immediately evaluate the effective mass
using the results in Tables \ref{table:fermi_liq_params_para} and
\ref{table:fermi_liq_params_ferro}.

In order to test the validity of our results, we compare the effective masses
obtained using Eqs.\ (\ref{eq:qem_flp_eval_para}) and
(\ref{eq:qem_flp_eval_ferro}) with the effective masses extracted directly
from the energy bands (reported in Ref.\ \onlinecite{ndd_effmass2}) in Table
\ref{table:comp_qem}.  The two measures of the effective mass disagree by a
statistically significant margin in half the cases.  The enormous uncertainty
in the finite-size extrapolation of the Fermi liquid parameters is the most
likely reason for the disagreement.  Direct calculation of the effective mass
using a fit to the energy band together with Eq.\ (\ref{eq:qem_eval}) is
likely to be more reliable, and we therefore suggest that the values of
$F_l^{a,s}/m^\ast$ given in Tables \ref{table:fermi_liq_params_para} and
\ref{table:fermi_liq_params_ferro} be multiplied by the effective mass
$m^\ast$ reported in Ref.\ \onlinecite{ndd_effmass2} to obtain the Fermi
liquid parameters.

\begin{table}
\begin{center}
\caption{Quasiparticle effective masses in a.u.\ for different density
  parameters $r_s$ and spin-polarization values $\zeta$, obtained directly
  from the energy band\cite{ndd_effmass2} and from the $F_1^s$ or $F_1$ Fermi
  liquid parameter [Eqs.\ (\ref{eq:qem_flp_eval_para}) and
    (\ref{eq:qem_flp_eval_ferro})]. \label{table:comp_qem}}
\begin{tabular}{ccr@{.}lr@{.}l} \\ \hline \hline

& & \multicolumn{4}{c}{Effective mass $m^\ast$} \\

\raisebox{1.5ex}[0pt]{$r_s$} & \raisebox{1.5ex}[0pt]{$\zeta$} &
\multicolumn{2}{c}{Ref.\ \onlinecite{ndd_effmass2}} &
\multicolumn{2}{c}{Eqs.\ (\ref{eq:qem_flp_eval_para}) and
  (\ref{eq:qem_flp_eval_ferro})} \\

\hline

$1$ & 0 & ~~~~$0$&$947(3)$ & ~~~~~~$0$&$85(2)$ \\

$5$ & 0 & $0$&$97(3)$ & $0$&$97(2)$ \\

$10$ & 0 & $0$&$85(6)$ & $0$&$9(2)$ \\

$1$ & 1 & ~~~~$0$&$841(3)$ & $0$&$82(2)$ \\

$5$ & 1 & $0$&$73(2)$  & $0$&$62(2)$ \\

$10$ & 1 & $0$&$67(4)$ & $0$&$493(7)$ \\

\hline \hline
\end{tabular}
\end{center}
\end{table}

\subsubsection{Isothermal compressibility}

The isothermal compressibility $\kappa^\ast$ of the interacting 2D HEG at zero
temperature satisfies
\begin{equation} \frac{\kappa}{\kappa^\ast} = \frac{r_s^4}{4(1+\zeta^2)} \left[
  \frac{\partial^2}{\partial r_s^2}-\frac{1}{r_s}\frac{\partial}{\partial r_s}
  \right]E(r_s,\zeta), \label{eq:comp_defn} \end{equation} where
$E(r_s,\zeta)$ is the total energy per electron as a function of density
parameter $r_s$ and spin polarization $\zeta$, and $\kappa$ is the isothermal
compressibility of the noninteracting system.\cite{giuliani}

A parameterization of the correlation energy per electron in paramagnetic 2D
electron gases is given in Ref.\ \onlinecite{ndd_2dheg_expvals}, so that we
may evaluate $\kappa/\kappa^\ast$ directly using Eq.\ (\ref{eq:comp_defn}).
We refer to this as the \textit{total-energy approach}.

Within Fermi liquid theory we have\cite{giuliani}
\begin{equation} \frac{\kappa}{\kappa^\ast} =
  \frac{1}{m^\ast}+\frac{F_0^s}{m^\ast}, \label{eq:comp_FL} \end{equation} for
a paramagnetic HEG and
\begin{equation} \frac{\kappa}{\kappa^\ast} =
  \frac{1}{m^\ast}+\frac{F_0}{m^\ast}, \label{eq:comp_FL_ferro} \end{equation}
for a ferromagnetic HEG, giving us a second approach for calculating the
isothermal compressibility, which we refer to as the \textit{Fermi-liquid
  approach}.  The value of $F_0^s/m^\ast$ is taken from Table
\ref{table:fermi_liq_params_para}, while the value of $m^\ast$ is taken from
Ref.\ \onlinecite{ndd_effmass2}.

A comparison of the isothermal compressibilities obtained using these two
different approaches is given in Table \ref{table:comp_comparison}.
Unfortunately the results are quite different.  We verified that the
compressibility ratios evaluated using the total-energy approach with two
different parameterizations of the correlation
energy\cite{ndd_2dheg_expvals,attaccalite_2002} agree to at least three
significant figures.  We therefore believe that the isothermal
compressibilities obtained from fits to ground-state DMC energy calculations
are reliable.

\begin{table}
\caption{Modification to the isothermal compressibility and spin
  susceptibility of a paramagnetic 2D HEG due to electron interactions as
  calculated (i) from Eqs.\ (\ref{eq:comp_defn}) and (\ref{eq:ss_ratio_defn})
  together with parameterizations of the total energy per
  particle\cite{ndd_2dheg_expvals,attaccalite_2002} and (ii) from
  Eqs.\ (\ref{eq:comp_FL}) and (\ref{eq:ss_ratio_FL}) together with the
  present calculation of the Fermi liquid properties.
  \label{table:comp_comparison}}
\begin{tabular}{cr@{.}lr@{.}lr@{.}lr@{.}l}
\hline \hline

 & \multicolumn{4}{c}{Compress.\ ratio $\kappa/\kappa^\ast$} &
\multicolumn{4}{c}{Spin-suscept.\ ratio $\chi/\chi^\ast$} \\

\raisebox{1.5ex}[0pt]{$r_s$} & \multicolumn{2}{c}{Tot.\ en.\ ap.} &
\multicolumn{2}{c}{Fermi liq.\ ap.} & \multicolumn{2}{c}{Tot.\ en.\ ap.} &
\multicolumn{2}{c}{Fermi liq.\ ap.} \\

\hline

1  & ~~~~~$0$&$533$ & ~~~$0$&$9(2)$ & ~~~~~$0$&$691$ & ~~~~~$0$&$89(8)$ \\

5  & $-1$&$735$ & $1$&$7(8)$ & $0$&$296$ & $3$&$0(7)$ \\

10 & $-4$&$989$ & $0$&$2(2)$ & $0$&$153$ & $4$&$3(7)$ \\

\hline \hline
\end{tabular}
\end{table}

The values of $F_0^s/m^\ast$ implied by the ground-state total-energy results
of Ref.\ \onlinecite{ndd_2dheg_expvals} together with the effective-mass data
reported in Ref.\ \onlinecite{ndd_effmass2} are
$F_0^s/m^\ast=\kappa/\kappa^\ast-1/m^\ast=-0.523(2)$, $-2.77(2)$, and
$-6.17(4)$ at $r_s=1$, $5$, and $10$, respectively.  These are relatively
close to the values of $F_0^s/m^\ast$ obtained at finite system sizes (see
Figs.\ \ref{fig:F_extrap_ert_N_rs1}, \ref{fig:F_extrap_ert_N_rs5}, and
\ref{fig:F_extrap_ert_N_rs10}).

The analogous results for a ferromagnetic HEG are shown in Table
\ref{table:comp_comparison_ferro}.  Again we see a significant difference
between the compressibilities obtained directly from the total energy and from
the Fermi liquid parameters.

\begin{table}
\caption{Modification to the isothermal compressibility of a ferromagnetic 2D
  HEG due to electron interactions as calculated (i) from
  Eqs.\ (\ref{eq:comp_defn}) together with a parameterization of the total
  energy per particle\cite{attaccalite_2002} and (ii) from
  Eqs.\ (\ref{eq:comp_FL_ferro}) together with the present calculation of the
  Fermi liquid properties.
  \label{table:comp_comparison_ferro}}
\begin{tabular}{cr@{.}lr@{.}l}
\hline \hline

 & \multicolumn{4}{c}{Compress.\ ratio $\kappa/\kappa^\ast$} \\

\raisebox{1.5ex}[0pt]{$r_s$} & \multicolumn{2}{c}{Tot.\ en.\ ap.} &
\multicolumn{2}{c}{Fermi liq.\ ap.} \\

\hline

1  & ~~~~~$0$&$680$ & ~~~$0$&$69(4)$ \\

5  & $-0$&$636$     & $-0$&$1(2)$ \\

10 & $-2$&$347$     & $-0$&$3(3)$ \\

\hline \hline
\end{tabular}
\end{table}

\subsubsection{Isothermal spin susceptibility}

The isothermal spin susceptibility $\chi^\ast$ of a paramagnetic HEG at zero
temperature satisfies\cite{giuliani}
\begin{equation} \frac{\chi}{\chi^\ast} = r_s^2 \left(\frac{\partial^2
    E}{\partial \zeta^2}
  \right)_{\zeta=0}, \label{eq:ss_ratio_defn} \end{equation} where $\chi$ is
the (Pauli) spin susceptibility of a free electron gas.  Attaccalite
\textit{et al.}\cite{attaccalite_2002}\ have reported a parameterization of
the correlation energy obtained in QMC calculations as a function of both
density parameter $r_s$ and spin polarization $\zeta$.  Hence we can use
Eq.\ (\ref{eq:ss_ratio_defn}) to evaluate $\chi/\chi^\ast$ using the
total-energy approach.

Within Fermi liquid theory the isothermal spin susceptibility $\chi^\ast$ of
an interacting electron system satisfies\cite{giuliani}
\begin{equation} \frac{\chi}{\chi^\ast} =
  \frac{1}{m^\ast}+\frac{F_0^a}{m^\ast}. \label{eq:ss_ratio_FL} \end{equation}
The value of $F_0^a/m^\ast$ is taken from Table
\ref{table:fermi_liq_params_para}, while the value of $m^\ast$ is taken from
Ref.\ \onlinecite{ndd_effmass2}.

We compare the isothermal spin susceptibilities obtained using these two
approaches in Table \ref{table:comp_comparison}.  The results obtained from a
fit to the ground-state energy as a function of density parameter and spin
polarization are quite different to the results obtained using our Fermi
liquid parameters.

The values of $F_0^a/m^\ast$ implied by the ground-state total-energy results
of Ref.\ \onlinecite{attaccalite_2002} together with the effective-mass data
reported in Ref.\ \onlinecite{ndd_effmass2} are
$F_0^a/m^\ast=\chi/\chi^\ast-1/m^\ast=-0.365(2)$, $-0.73(2)$, and $-1.02(4)$
at $r_s=1$, $5$, and $10$, respectively.  These are relatively close to the
results obtained at finite system size shown in
Figs.\ \ref{fig:F_extrap_ert_N_rs1}, \ref{fig:F_extrap_ert_N_rs5}, and
\ref{fig:F_extrap_ert_N_rs10}.  The finite-size extrapolation
appears to move the Fermi liquid parameters away from the values suggested by
the spin susceptibility.

Both the isothermal compressibility and spin susceptibility results show that
we are not able to extrapolate the Fermi liquid parameters to the
thermodynamic limit with quantitative accuracy.  There is no clear numerical
evidence to support the $N^{-1/4}$ scaling, and in most cases any systematic
finite-size error appears to be swamped by oscillations due to shell-filling
effects in the Fermi liquid parameters as a function of system size.

\subsection{Summary of the ``best available'' Fermi liquid parameters}

In Tables \ref{table:fermi_liq_params_best_para} and
\ref{table:fermi_liq_params_best_ferro} we summarize the Fermi liquid
parameters determined from QMC results reported in
Refs.\ \onlinecite{ndd_effmass2,ndd_2dheg_expvals,attaccalite_2002}.  The
values of $F_1^s$ and $F_1$ are determined using the effective masses reported
in Ref.\ \onlinecite{ndd_effmass2}, the values of $F_0^s$ are determined using
the effective masses of Ref.\ \onlinecite{ndd_effmass2} together with the
parameterization of the correlation energy given in
Ref.\ \onlinecite{ndd_2dheg_expvals}, and the values of $F_0$ and $F_0^a$ are
determined using the effective masses together with the parameterization of
the correlation energy given in Ref.\ \onlinecite{attaccalite_2002}.

\begin{table*}
\begin{center}
  \caption{Best available Fermi liquid parameters of the paramagnetic 2D HEG
    as inferred from QMC results in the literature (see text).
\label{table:fermi_liq_params_best_para}}
\begin{tabular}{cr@{}lr@{}lr@{}l} \\ \hline \hline

$r_s$ & \multicolumn{2}{c}{$F_0^s$} & \multicolumn{2}{c}{$F_0^a$} &
  \multicolumn{2}{c}{$F_1^s$} \\

\hline

$1$  & $-0$&$.495(2)$ & $-0$&$.346(2)$ & $-0$&$.053(3)$ \\

$5$  & $-2$&$.68(5)$  & $-0$&$.713(9)$ & $-0$&$.03(3)$ \\

$10$ & $-5$&$.2(3)$   & $-0$&$.870(9)$ & $-0$&$.15(6)$ \\

\hline\hline
\end{tabular}
\end{center}
\end{table*}

\begin{table}
\begin{center}
\caption{Best available Fermi liquid parameters of the ferromagnetic 2D HEG as
  inferred from QMC results in the literature (see text).
\label{table:fermi_liq_params_best_ferro}}
\begin{tabular}{cr@{}lr@{}l} \\ \hline \hline

$r_s$ & \multicolumn{2}{c}{$F_0$} & \multicolumn{2}{c}{$F_1$} \\

\hline

$1$ & $-0$&$.428(2)$ & $-0$&$.159(3)$ \\

$5$ & $-1$&$.46(1)$ & $-0$&$.27(2)$ \\

$10$ & $-2$&$.57(9)$ & $-0$&$.33(4)$ \\

\hline\hline
\end{tabular}
\end{center}
\end{table}

\section{Conclusions \label{sec:conclusions}}

We have used QMC methods to calculate the Fermi liquid parameters of the 2D
HEG\@.  However, the results we have obtained are inconsistent with more
direct evaluations of the isothermal compressibility and spin
susceptibility. Determining the Fermi liquid parameters therefore remains a
grand-challenge problem due to the enormous difficulty in extrapolating the
QMC data to the thermodynamic limit.  Nevertheless, we have been able to
describe the difficulties of determining the Fermi liquid parameters using QMC
techniques, and we have assembled a set of ``best available'' values of some
of the parameters.  Our work demonstrates considerable progress in determining
accurate values for the Fermi liquid parameters of the 2D HEG\@.  Although the
quasiparticle effective masses deduced from our determination of the Fermi
liquid parameters are only in approximate agreement with the values obtained
directly from the energy band,\cite{ndd_effmass2} neither method shows mass
enhancement at low densities.

\begin{acknowledgments}
Financial support was received from Lancaster University under the Early
Career Small Grant Scheme, and the Engineering and Physical Sciences Research
Council.  This research used resources of the Oak Ridge Leadership Computing
Facility at the Oak Ridge National Laboratory, which is supported by the
Office of Science of the U.S.\ Department of Energy under Contract
No.\ DE-AC05-00OR22725.  Additional computing resources were provided by the
Cambridge High Performance Computing Service.
\end{acknowledgments}

\end{document}